\documentclass[epj]{svjour}
\usepackage{graphics}
\usepackage{cuted}   
\usepackage{graphicx}
\usepackage{hyperref}
\usepackage{amsmath,amssymb,amsfonts}
\usepackage{mathrsfs}
\usepackage{enumitem}

\usepackage{cite}
\usepackage{csquotes}
\begin{document}

\title{Relativistic compact object in Generalised Tolman - Kuchowicz spacetime with quadratic equation of state}

\author{Hemani R. Acharya\inst{1} \and D. M. Pandya\inst{1} \and Bharat Parekh\inst{1} \and V. O. Thomas\inst{2}}

\institute{
\inst{1}Department of Physics, Pandit Deendayal Energy University, Knowledge Corridor, Gandhinagar 382426, Gujarat, India\\
\email{acharya.hemani@gmail.com}
\and
\inst{1}Department of Mathematics, Pandit Deendayal Energy University, Gandhinagar 382426, Gujarat, India\\
\email{dishantpandya777@gmail.com}
\and
\inst{1}Department of Physics, Pandit Deendayal Energy University, Knowledge Corridor, Gandhinagar 382426, Gujarat, India\\
\email{bharat.parekh@sot.pdpu.ac.in}
\and
\inst{2}Department of Mathematics, Faculty of Science, The M. S. University of Baroda 390002, Vadodara, Gujarat, India\\
\email{votmsu@gmail.com}
}

\abstract{This paper presents the class of solutions to the Einstein field equations for the uncharged static spherically symmetric compact object {PSR J0952\textendash0607} by using Generalized Tolman - Kuchowicz space-time metric with quadratic equation of state. We have obtained the bound on the model parameter n graphically and achieved the stable stellar structure of the mathematical model of a compact object. The stability of the generated model is examined by the Tolman - Oppenheimer - Volkoff equation and the Harrison-Zeldovich-Novikov criterion. This anisotropic compact star model fulfills all the required stability criteria including the causality condition, adiabatic index, and Buchdahl condition, Herrera's cracking condition and pertains free from central singularities.}
\PACS{04.20.-q~\and~04.20.Jb~\and~04.40.Dg~\and~12.39.Ba}
\authorrunning{Acharya et al.}
\titlerunning{Generalised Tolman - Kuchowicz spacetime with quadratic equation of state}
\maketitle

\section{Introduction}\label{sec1}
 Einstein's General Relativity (GR) has continuously proven its accuracy over time by predicting various astrophysical phenomena. This theory has been pivotal from explaining the orbital precession of the Mercury to the more recent success in the detection of gravitational waves during the collision of black holes confirming its applicability in astrophysics and cosmology \cite{Einstein1916}. Stars form within the clouds of dust and gas distributed unevenly across most galaxies. During the later stages of their life cycle, heavy stars reach a stage where the outward pressure generated by nuclear fusion is insufficient to balance the inward pull of gravitational force. At this stage, stars collapse under their own gravity, leading to what is termed as stellar death \cite{sagert2006compact}. The collapse of a star leads to the formation of a compact star like white dwarf, neutron star, quark star and black hole depending on the initial mass of the star. Compact stars are in the relativistic regime and hence such objects are to be investigated with the help of Einstein's field equations (EFEs). Since EFEs form a system of highly non-linear partial differential equations, it is often difficult to obtain closed form solutions. However, a number of physically significant solutions are available at present. Schwarzschild provided the first interior \cite{Schwarzschild1916}  and exterior solutions \cite{Schwarzschild1916_ExactSolution} for a spherically symmetric distribution matter with uniform density. Out of the 127 solutions of EFEs studied by Delgaty and Lake \cite{delgaty1998physical}, for physical plausibility only 16 of them met all criteria and only 9 solutions exhibited a decreasing sound speed with radius.
 	\subsection{Anisotropy and its effect on Compact star}\label{1.1}
 
 The compact stars are generally considered to be isotropic and spherically symmetric. However, due to their higher densities and strong gravitational fields, which create tremendous internal pressures, they often lead to deviations from this situation. These deviations showcase as anisotropic pressures, categorized into radial $p_r$ and transverse $p_\perp$ pressures, which are mutually perpendicular. This yields an anisotropic factor ($\Delta = p_\perp - p_r$) which analyses how the internal structure of these objects deviate from the isotropic case \cite{kumar2022relativistic}. The possibility of a distinction between transverse and radial pressures in stars was proposed by Lemaitre \cite{lemaitre1933univers}.  Ruderman later, introduced the concept of anisotropy originally \cite{ruderman1972pulsars}. It is believed that anisotropy forms in compact stars in regions where densities exceed $10^{15} g/cc$. Ruderman and Canuto \cite{ruderman1972pulsars, Canuto1973} have observed that the treatment of nuclear interactions becomes relativistic under such circumstances. Anisotropy in stars can be caused by superfluid neutrons  \cite{kippenhahn1990stellar}, a solid core, pion condensation phenomena \cite{sawyer1972condensed}, phase transitions \cite{sokolov1980phase}, effects of slow rotation \cite{herrera1995jeans}, strong magnetic fields \cite{weber2017pulsars}, and the blending of two distinct fluids \cite{letelier1980anisotropic, bayin1982anisotropic}. Works of Herrera and Santos \cite{herrera1997local} and Chan et al. \cite{chan2003mfa} provided an understanding of the physical phenomena producing the pressure anisotropy and analysed the significance of local anisotropy. Dev and Gleiser \cite{dev2002anisotropic, dev2003anisotropic} examined multiple factors that influence pressure anisotropy.
 According to Herrera \cite{herrera2020stability}, \enquote{Even if the system is originally thought to be isotropic, physical processes expected in star evolution tend to generate pressure anisotropy}. Therefore, while working with relativistic fluids, anisotropy must be taken into consideration.
 
 \subsection{Equation of State (EoS)}\label{1.3}
 The physical structure, nuclear density, and material composition inside the star remain unresolved. At extremely high densities, the composition of matter inside compact stars likely transforms into such states that may involve hyperons, quark matter, or other forms of matter beyond nuclear densities. Such complexity demands the use of advanced theoretical frameworks. To address this problem, one of the approaches is to choose a particular Equation of State (EoS) which represents a relation between radial pressure and energy density, $p_r = p_r(\rho)$ \cite{kumar2024exploring}. The EoS is essential for several astronomical phenomena that are characterised by very high nuclear densities and temperatures, such as (i) neutron star mergers in binary systems (ii) cold Neutron Stars (NS) or Black Holes (BH) as well as NS-NS and NS-BH mergers \cite{burgio2021neutron}. EoS has an impact on the compact star's dynamic structure and nucleosynthesis process. According to the theory of weak interaction, when the density increases, some of the nucleons can convert into hyperons if the fermi energy surpasses the hyperons mass. This action lowers the Fermi pressure and results in a softening of the EoS \cite{xu2010isospin}. Recent studies suggest that the interaction among electrons, nucleons, hyperons, and quark matter inside compact stars may support a maximum mass of up to 
 $2 M_\odot$ \cite{weissenborn2011quark, bednarek2012hyperons}. Neutron- neutron interaction in the highly dense neutron star enables the emergence of cooper pairs at an extreme temperature at $10^{10}$K leading to the emergence of superfluidity \cite{migdal1959superfluidity}. In their studies, Friedman and Pandharipande \cite{friedman1981hot} as well as Wiringa et al. \cite{wiringa1988equation} adopted an Equation of State (EoS) describing the ground-state composition of $npe\mu$-matter, incorporating realistic two and three-body nucleon interactions. This approach results in a stiffer EoS, predicting a maximum mass in the range of approximately $1.9$--$2.1~M_{\odot}$.
 With the help of the above studies, it is known that a stiffer EoS can sustain maximum masses of roughly $1.9–2.1 M_{\odot}$. It is postulated by Migdal \cite{migdal1972stability, migdal1972phase} that superconducting pions might arise in the compact star due to $\pi - n$ strong interaction and the very stiff EoS that containing superconducting pions can support the mass upto 3 $M_\odot$ \cite{mao2014structure}. Many researchers make the mathematical model for the relativistic star using Quadratic EoS (QEoS) \cite{feroze2011charged, maharaj2012regular, sharma2013relativistic, ngubelanga2015compact, bhar2016compact, bhar2017compact, malaver2018some, sunzu2018new, malaver2020relativistic, kumar2024exploring, pant2024comprehensive, malaver2014strange}.
 In 1939, Tolman \cite{tolman1939static} independently provided eight different types of metric potential to solve the EFEs. In 1968, Kuchowicz \cite{kuchowicz1968general} introduced metric potential which is free from the singularity to get the compact structure. Jasim et al. \cite{jasim2018anisotropic} provided the solution of a strange star by using the Tolman-Kuchowicz (T-K) metric. In 2019, Biswas et al. \cite{biswas2019relativistic} presented the T-K metric with MIT bag model EoS. In 2019, Maurya and Tello-Ortiz \cite{maurya2019charged} developed a mathematical model for the charged stars in the T-K spacetime. In 2021, Rej et al \cite{rej2021charged} presented the charged configuration of a compact star with MIT bag model under the f(R,T) modified gravity. In 2022, Bhar \cite{bhar2021dark} studied dark energy stars in T-K spacetime within Einstein gravity, analyzing their stability and physical properties. Using the T-K spacetime, Rej and Bhar in 2021 investigated hybrid stars containing baryonic and strange quark matter.  Rej and  Karmakar (2023) \cite{rej2023charged} investigated the stability and physical properties of charged strange stars with anisotropic dark energy by representing stars in T-K spacetime. Using the Tolman - Kuchowicz metric, Bhar in 2023 \cite{bhar2023compact} developed a model of anisotropic compact stars in f(T) gravity and examined its stability and physical acceptability. In 2019, The stability and physical characteristics of anisotropic compact stars in T–K spacetime under five-dimensional Einstein–Gauss–Bonnet gravity have been studied by Bhar et al. \cite{bhar2019compact}. Anisotropic strange stars under f(R,T) gravity have been modeled in T–K spacetime by Biswas et al. \cite{biswas2020anisotropic}. Also, their stability and physical feasibility have been examined.
 To investigate the charged stellar structure within the context of Tolman-Kuchowicz spacetime, Shamir and Fayyaz \cite{farasat2020charged} have solved Einstein-Maxwell field equations and examined the physical characteristics of compact stars under the f(R) gravity model. Recently in 2023, Das et al. \cite{das2023models} presented the construction of a mathematical model of a compact star in the framework of the Generalized Tolman - Kuchowicz (GTK) metric. \\
 Despite significant efforts put in this field, there is still a need to work the GTK metric with the QEoS to analyse the properties of the star up to the mass above $2 M_\odot$. Therefore, the purpose of this study is to construct the mathematical model of uncharged compact star under the framework of GTK with the help of QEoS. \\ 	
 To accomplish the aforementioned goal, section \ref{sec2} presents the fundamental Einstein field equations. Section \ref{sec3} provides the solution of the EFEs utilising QEoS.  The interior and exterior metrics used to obtain the model parameters as described in section \ref{sec4}. The physical acceptability conditions for making the realistic model of the compact star are presented in section \ref{sec5}.  Section \ref{sec6} presents the physical analysis through both graphical and analytical methods. The conclusion of the study is provided in section \ref{sec7}.

\section{Basic Mathematical formulation of Einstein field Equations}\label{sec2}
The line element of the spherically symmetric space-time metric to explain the interior geometry of a super-dense star is represented in the standard co-ordinates  $ x^\mu $ = (t, r, $\theta$, $\phi$) as, 
\begin{equation}
	ds^2= e^\nu dt^2- e^\lambda dr^2- r^2(d\theta^2 + sin^2\theta  d\phi^2) , \label{equ:1}
\end{equation}
where $\nu(r)$ and $\lambda(r) $  are the temporal and spatial metric coefficients that depend only on the radial coordinate. The GTK  metric is described by the ansatz, 
\begin{equation}
	e^\lambda = (1+ ar^2+ br^4)^n \\ \space , ~ \space e^\nu = C^2 e^{Ar^2}  \label{equ:2}
\end{equation}
Here a,~ b,~ A, and C are the arbitrary constants that can be evaluated based on the appropriate boundary conditions. The constant n is the positive number (n $\geq$ 1). The modified metric potential $g_{tt}$ in the GTK \cite{das2023models} ansatz has an exponent n and reduces to the original T-K ansatz for n = 1 (Tolman 1939 \cite{tolman1939static}; Kuchowicz 1968 \cite{kuchowicz1968general}).\\
The Einstein field equations are given by: 
\begin{equation}
	R_{ij} - \frac{1}{2}R\space g_{ij} = 8\pi T_{ij}                           \label{equ:3}
\end{equation}
where \( R_{ij} \), \( R \), \( T_{ij} \), and \( g_{ij} \) are the Ricci tensor, Ricci scalar, energy-momentum tensor and metric tensor of the fluid distribution, respectively. In geometric units (\( G = C^2 = 1 \)), an anisotropic imperfect fluid has the following energy-momentum tensor: 
\begin{equation}
	T_{ij} = (\rho+ P_{\perp})u_iu_j + P_{\perp}\space g_{ij}+ (P_r-P_{\perp})\chi_i \chi_j \label{equ:4}
\end{equation}
where $\rho, P_{\perp}$ and $P_{r}$ are energy density, transverse, and radial pressure, respectively. The fluid's unit four-velocity and space-like vector components are denoted as  $u_i$ and $\chi_i$ respectively. \\
The EFEs are represented by the following set of three equations,
\begin{equation}
	8\pi\rho = \frac{1}{r^2}-e^{-\lambda}(\frac{1}{r^2}-\frac{\lambda'}{r}),	\label{equ:5}
\end{equation}
\begin{equation}
	8\pi p_r = e^{-\lambda} (\frac{1}{r^2}+ \frac{\nu'}{r})-\frac{1}{r^2}, 	\label{equ:6}
\end{equation}
\begin{equation}
	8\pi p_\perp = \frac{e^{-\lambda}}{4}[2\nu''+ (\nu'- \lambda')(\nu' + \frac{2}{r})]. 	\label{equ:7}
\end{equation}
The differentiation with respect to r is represented by a prime. 

Further modification of equations (\ref{equ:5}) - (\ref{equ:7}) can take the following form \cite{gokhroo1994anisotropic}:
\begin{equation}
	e^{-\lambda} = 1 - \frac{2m}{r}, \label{equ:8}
\end{equation}
\begin{equation}
	(1 - \frac{2m}{r})\nu' = 8 \pi p_r + \frac{2m}{r^2}, \label{equ:9}
\end{equation}
\begin{equation}
	-\frac{4}{r}(8\pi\sqrt3 S)= (8 \pi \rho + 8 \pi p_r)\nu' + 2(8\pi p_r') , \label{equ:10}
\end{equation}
Here,
\begin{equation}
	m(r) = 4 \pi \int_0^r u^2 \rho (u) du.  \label{equ:11}
\end{equation}
The radial pressure $p_r$ and the matter density $\rho$ are assumed to be related by the QEoS as follows:
\begin{equation}
	p_r = \alpha \rho^2 + \beta \rho - \gamma,  \label{equ:12}
\end{equation}
where $\alpha,\space\beta$ and $\gamma$ are arbitrary real constants.
\section{Solution of the Field Equations in GTK metric}\label{sec3}

By using the GTK metric equation (\ref{equ:2}) and the quadratic EoS equation (\ref{equ:12}), the EFEs (\ref{equ:5}) - (\ref{equ:7}) are expressed as follows: 

\begin{equation}
	\rho = \frac{1}{r^2} + 2n \left(a + 2b r^2\right) \left(1 + a r^2 + b r^4\right)^{-1 - n} - \frac{\left(1 + a r^2 + b r^4\right)^{-n}}{r^2}, \label{equ:13}
\end{equation}

\begin{equation}
	p_r = \left( \frac{X}{r} + \frac{Y}{r^2} \right)^2 \alpha + \left( \frac{X}{r} + \frac{Y}{r^2} \right) \beta - \gamma, \label{equ:14}
\end{equation}
where, 
\begin{align}
	X = n \left( 2 a r + 4 b r^3 \right) \left( 1 + a r^2 + b r^4 \right)^{-1 - n}, \nonumber \\ 
	Y = 1 - \left( 1 + a r^2 + b r^4 \right)^{-n}.\nonumber
\end{align}
Differentiating equation (\ref{equ:14}) with respect to r, 

\begin{align}
	8 \pi \frac{dp_r}{dr} 
	&= \frac{(-1 - 2 j + s)\alpha}{r^4} 
	\Bigl[\, 2(-1-n)n r^2 w f q^{-2-n} - 8 b n r^3 q^{-1-n} \notag \\[5pt]
	&\quad - 4 n r w q^{-1-n} - n f q^{-1-n}\,\Bigr] 
	- \frac{4(-1 - 2 j + s)^2 \alpha}{r^5} \notag \\[5pt]
	&\quad + \frac{\beta}{r^2}\Bigl[\,2(-1-n)n r^2 w f q^{-2-n} + 8 b n r^3 q^{-1-n} \notag \\[5pt]
	&\quad + 4 n r w q^{-1-n} + n f q^{-1-n}\,\Bigr] 
	- \frac{2(1 + 2 j - s)\beta}{r^3}.
	\label{equ:15}
\end{align}

\[
\begin{aligned}
	q = 1 + a r^2 + b r^4, 
	w &= a + 2 b r^2,
	f = 2 a r + 4 b r^3,\\
	j &= n r^2 w q^{-1-n},
	s = q^{-n}.
\end{aligned}
\]

By using equations (\ref{equ:12}), (\ref{equ:13}), and (\ref{equ:14}), equation (\ref{equ:10}) can be represented as follows:\\

	\begin{multline}\label{equ:16}
	8 \pi \sqrt{3} S = \frac{q^{-3 - 2n}}{r^4} \biggl[ -1 + h + 2 b r^4 (-1 - 3n + h) \\
	+ a^2 r^4 (-1 + n + 2 n^2 + h) + b^2 r^8 (-1 + 2n + 8 n^2 + h) \\
	+ a r^2 \left(8 b n^2 r^4 + n (-1 + b r^4) + 2 (1 + b r^4) (-1 + h)\right) \biggr] \\
	\times \biggl[2 \left(-1 + h + a r^2 (-1 + 2n + h) + b r^4 (-1 + 4n + h)\right)\alpha \\
	+ r^2 q^{1 + n} \beta \biggr] \\
	- \frac{A}{2 r^2}\biggl[2 n r^4 w z + r^2 (1 - s) + (-1 - 2 n r^2 w z + s)^2 \alpha \\
	+ r^2 (1 + 2 n r^2 w z - s) \beta - r^4 \gamma\biggr].
\end{multline}

\text{where}, \\
\[
\ h = q^n, \ z = q^{-1-n}.
\]

At r = 0, anisotropy is zero. The form of the equation $8 \pi p_\perp = 8 \pi p_r - 8 \pi \sqrt3 S$ is as follows: \\

\begin{multline}\label{equ:17}
	P_{\perp} = \frac{\alpha}{r^4} \left(-1 - 2 n r^2 w z + s\right)^2 
	+ \frac{\beta}{r^2} \left(1 + 2 n r^2 w z - s\right) \\
	- \frac{q^{-3 - 2n}}{r^4}\biggl[-1 + h + 2 b r^4 (-1 - 3 n + h) \\
	+ a^2 r^4 (-1 + n + 2 n^2 + h) 
	+ b^2 r^8 (-1 + 2 n + 8 n^2 + h) \\
	+ a r^2 \left(8 b n^2 r^4 + n (-1 + b r^4) + 2 (1 + b r^4)(-1 + h)\right)\biggr] \\
	\times\biggl[2 \left(-1 + h + a r^2 (-1 + 2 n + h) + b r^4 (-1 + 4 n + h)\right)\alpha \\
	+ r^2 q^{1 + n}\beta\biggr] \\
	- \gamma + \frac{A}{2 r^2}\biggl[2 n r^4 w z + r^2(1 - s) \\
	+ \left(-1 - 2 n r^2 w z + s\right)^2\alpha \\
	+ r^2 \left(1 + 2 n r^2 w z - s\right)\beta - r^4\gamma\biggr].
\end{multline}

\section{Boundary conditions on the model parameters}\label{sec4}
The central density can be calculated by considering r = 0 in equation (\ref{equ:13}),\\

\begin{equation} 
	\rho_c = 2 a n. \space \label{equ:18}
\end{equation}
From equations (\ref{equ:14}) and (\ref{equ:17}), we notice that 
\begin{equation}
	p_r(0) = p_\perp(0). \space \label{equ:19}
\end{equation}
\par The matching conditions are utilised to determine the unknowns $a, ~ b,~ A$, and $C$ at the boundary of the compact star. At the boundary r = R, the GTK metric should match continuous with the Schwarzschild metric. 

	\begin{equation}
	ds^2 = \left( 1 - \frac{2M}{r} \right) dt^2 - \left( 1 - \frac{2M}{r} \right)^{-1} dr^2 - r^2 \left( d\theta^2 + \sin^2 \theta \, d\phi^2 \right), \quad \label{equ:20}
\end{equation}

where the mass of the star is denoted by M. At r = R (radius of the star), the metric coefficients $g_{tt}$, $g_{rr}$ and $\frac{\partial g_{tt}}{\partial r}$ must remain continuous through the interior and exterior metrics. \\These give the following set of three equations:\\
\begin{equation}
	g_{tt} : 1 -\frac{2M}{R} = C^2 e^{AR^2},  \label{equ:21}
\end{equation}
\begin{equation}
	g_{rr} : \left( 1 - \frac{2M}{R} \right)^{-1} = \left( 1 + a R^2 + b R^4 \right)^n, \quad \label{equ:22}
\end{equation}

\begin{equation}
	\frac{\partial g_{tt}}{\partial r}: \frac{M}{R} = A R^2 C^2 e^{AR^2}. \label{equ:23}
\end{equation}
At the boundary r = R, the radial pressure $P_r$ must vanish.
this gives,
\begin{equation}
	p_r(R)=0. \label{equ: 24}
\end{equation}
Equations (\ref{equ:21}) - (\ref{equ: 24}) determine the constants  $a, b, A$, and $C$ as follows:
\begin{multline}\label{equ:25}
	a = \frac{1}{4 R^2 \alpha n} \biggl[R^2 \beta u^{n + 1} - R^2 u \sqrt{u^{2n} (4 \alpha \gamma + \beta^2)} \\
	+ 8 \alpha n u - 8 \alpha n - 2 \alpha u + 2 \alpha u^{n + 1}\biggr].
\end{multline}

\begin{equation} 
	b = \frac{U - 1- a R^2}{R^4}, \label{equ:26}
\end{equation}
\begin{equation}
	A = \frac{M}{R^3 (1 - \frac{2M}{R})}, \label{equ:27}
\end{equation}
\begin{equation}
	C = \left( e^{- \frac{A R^2}{2}} \right) \left( 1 - \frac{2M}{R} \right)^{\frac{1}{2}}, \quad \label{equ:28}
\end{equation}

where $U = \left(1- \frac{2M}{R}\right)^ {-\frac{1}{n}}$.
\par The values of $a,~b,~A,~C$ can be obtained from equations (\ref{equ:25}) - (\ref{equ:28}) by choosing appropriate values for arbitrary constants $~\alpha,~\beta, ~\gamma$ and $n$. 

	\section{Fundamental criteria for physical acceptability conditions}\label{sec5}
\par To be a physically acceptable model, the interior solution of the gravitational field equations must adhere to some physical conditions\cite{kuchowicz1972differential, herrera1997local, abreu2007sound}.
\begin{enumerate} [label=\protect\textup{(\Roman*)}]
	\item The solution must be free from physical and geometric singularities, ensuring finite and positive central pressure and density, as well as positive, and nonzero values for $(e^\lambda)$ and $(e^\nu)$. equation (\ref{equ:2}) yields $(e^\lambda) = 1$ and $(e^\nu) = c^2$ at r = 0.
	
	\item The density $\rho$, both radial and transverse pressures $p_r$ and $p_\perp$,  respectively should remain positive, finite, and must decrease towards the exterior.
	
	\item At the center $(r =0)$, the radial and the transverse pressure must be equal, i.e., $p_r(0)=p_\perp(0)$, providing that the anisotropy at the center is zero, i.e., $\Delta(0)= 0$ \cite{bowers1974anisotropic,mccormick2014first}.
	
	\item The radial pressure is zero at the surface, i.e., $p_r(R)=0$, where R is the boundary of the star.
	
	\item Within the region $0\leq r \leq R$, the gradients of pressures and density should be non-positive, i.e., $\frac{dp_r}{dr},\frac{dp_\perp}{dr},\frac{d\rho}{dr} \leq 0$.
	
	\item To ensure a physically acceptable model, the causality conditions must be satisfied. These conditions are described by $0 \leq v^2_{r} (= \frac{dp_r}{d\rho}) \leq 1$ and $0 \leq v^2_{\perp} (= \frac{dp_\perp}{d\rho}) \leq 1$. where, $c =1$ is considered.
	
	\item The critical value $\Gamma_{\text{crit}}$ within the stellar interior should be exceeded by the adiabatic index $\Gamma$ \cite{bondi1964contraction}.
	
	\item All energy conditions, including the Null Energy Condition $(NEC)$, Weak Energy Condition $(WEC)$, Strong Energy Condition $(SEC)$, Dominant Energy Condition $(DEC)$, and Trace Energy Condition $(TEC)$, must be fulfilled inside the star.
\end{enumerate}

\section{Physical Analysis}\label{sec6}
The gravitational field equations for the interior of a static fluid sphere must satisfy the fundamental physical viability conditions to establish the physical validity of stellar configurations. Graphical and analytical methods are used to verify the physical validity of the proposed mathematical model. The study focuses on the recently observed neutron star, {PSR J0952\textendash0607}, which has an observed mass of \(2.35 \pm 0.17 \, M_{\odot}\). The minimum value of the maximum mass of NS is $2.19  M_{\odot}$ (at \(1\sigma\) confidence) and \(2.09 \, M_{\odot}\) (at \(3\sigma\) confidence), with an approximate radius of 10 km \cite{romani2022psr}. The physical characteristics of the compact object are examined within the framework of the GTK metric, and the model parameters for {PSR J0952\textendash0607}  are presented in Table\ref{tab:Table 1}.

\begin{table}[h!]
	\centering
	\footnotesize
	\caption{The values of parameters \( a \), \( b \), \( A \), and \( C \) for {PSR J0952\textendash0607} (\( M = 2.17 M_\odot \), \( R = 9.56 \, \mathrm{km} \), \( \alpha = 0.7 \), \( \beta = 0.06 \)).}
	\label{tab:Table 1}
	\resizebox{0.5\textwidth}{!}{
		\begin{tabular}{cccccc}
			\hline
			$n$    & $a$           & $b$            & $A$           & $C$        & $\gamma$ \\ \hline
			1      & 0.027785641   & -0.000059624   & 0.006600163   & 0.424666   & 0.000746 \\
			1.5    & 0.01734223    & 0.000054644    & 0.006600163   & 0.424666   & 0.000975 \\
			2      & 0.01300223    & 0.000102131    & 0.006600163   & 0.424666   & 0.0010   \\
			2.38   & 0.0107848     & 0.000126393    & 0.006600163   & 0.424666   & 0.00097  \\ \hline
	\end{tabular}}
\end{table}

	\subsection{Radial profile of density and pressures}

\par
\begin{figure}[h!]
	\centering
	\includegraphics[width=0.5\textwidth]{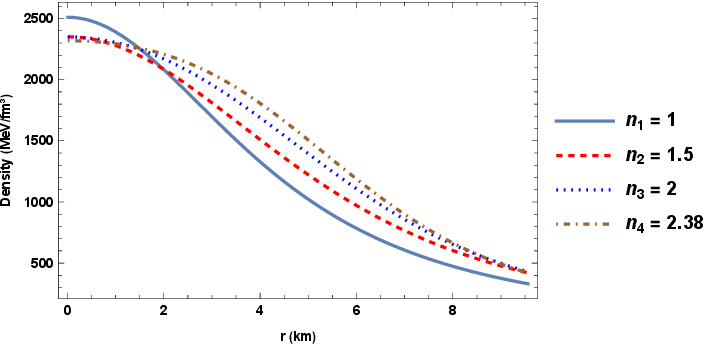} 
	\caption{
		\centering
		\textbf{Change in radial distribution of matter density} (\( \rho \)) \textbf{in {PSR J0952\textendash0607}} 
		\newline (\( M = 2.17 \, M_{\odot}; R = 9.56 \, \mathrm{km} \)) for various values of \( n \).
	}
	\label{fig:1}
\end{figure}

\begin{figure}[h!]
	\centering
	\includegraphics[width=0.5\textwidth]{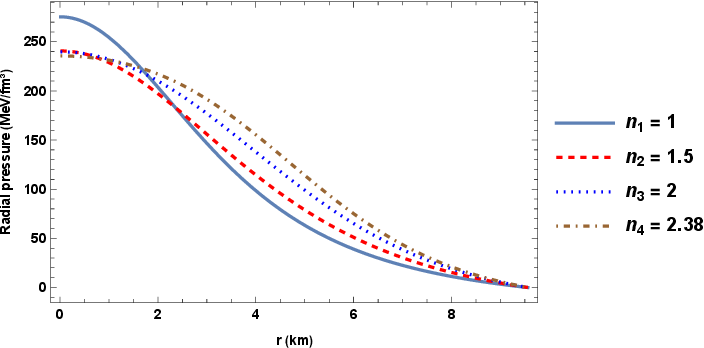} 
	\caption{
		\centering
		\textbf{Change in radial distribution of radial pressure} \( (P_r) \) \textbf{in {PSR J0952\textendash0607}} 
		\newline (\( M = 2.17 \, M_{\odot}; R = 9.56 \, \mathrm{km} \)) for various values of \( n \).
	}
	\label{fig:2}
\end{figure}

\begin{figure}[h!]
	\centering
	\includegraphics[width=0.5\textwidth]{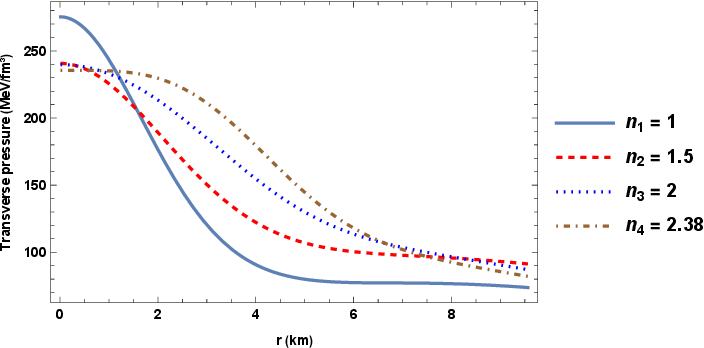} 
	\caption{
		\centering
		\textbf{Change in radial distribution of transverse pressure} \( (P_\perp) \) \textbf{in {PSR J0952\textendash0607}} 
		\newline (\( M = 2.17 \, M_{\odot}; R = 9.56 \, \mathrm{km} \)) for various values of \( n \).
	}
	\label{fig:3}
\end{figure}

 Figure \ref{fig:1}, \ref{fig:2}, and \ref{fig:3} illustrate the radial variations in energy density $\rho$, radial pressure $p_r$, and transverse pressure $p_{\perp}$ for different values of $n$ respectively. These parameters are positive throughout the star interior, reaching a maximum at the core and decreasing outward.

Furthermore, the central density ($\rho_c$) can be obtained as follows:
\begin{equation}
	\rho_c = 2 a n \label{equ:29}
\end{equation}

Similarly, central pressure ($p_c$) can be represented as follows:
\begin{equation}
	p_c = (2 a n)^2 \alpha + (2 a n) \beta - \gamma \label{equ:30}
\end{equation}

Based on Figure \ref{fig:1}, it has been observed that at the center of the star, energy density decreases as n increases. The variations of radial and transverse pressures as given in Figure \ref{fig:2} and Figure \ref{fig:3}, respectively, indicate that these pressures decrease as n increases. The transverse pressure has some non-zero value while the radial pressure vanishes at the boundary of the star. The difference between transverse and radial pressure gives the notion of anisotropy.  

\subsection{Anisotropy factor}
\begin{figure}[h!]
	\centering
	\includegraphics[width=0.5\textwidth]{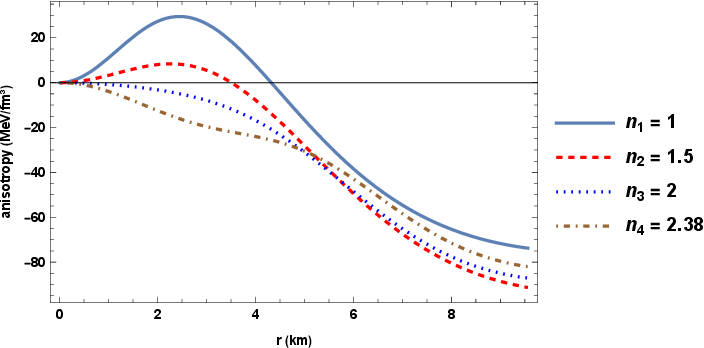} 
	\caption{
		\centering
		\textbf{Change in radial distribution of anisotropy factor} \( (S) \) \textbf{in {PSR J0952\textendash0607} } 
		\newline (\( M = 2.17 \, M_{\odot}; R = 9.56 \, \mathrm{km} \)) for various values of \( n \).
	}
	\label{fig:4}
\end{figure}

The radial variation of the anisotropy $S$ for various values of n is illustrated in Figure \ref{fig:4}. For $n_1 =1$ and $n_2=1.5$, $S$ is positive near origin, showing that the radial pressure $p_r$ is higher than the transverse pressure $p_\perp$, leading to an inward force. As the radius increases, the $S$ becomes negative, showing that $p_\perp$ exceeds $p_r$, leading to an outward force. For higher values of $n$ , i.e., $n_3 = 2$ and $n_4 = 2.38$, $S$ remains negative throughout, suggesting a consistent dominance of transverse pressure over radial pressure, which can influence the star's structural stability.

\subsection{Herrera cracking criteria}
The causality conditions are examined to make a realistic model of self-gravitating systems. For these conditions to hold, the square of the radial sound speed \((v_r^2 = \frac{dp_r}{d\rho})\) and the square of the transverse sound speed \((v_t^2 = \frac{dp_\perp}{d\rho})\) should be in the range between 0 and 1.(\cite{herrera1992cracking}, \cite{abreu2007sound}).

\begin{figure}[h!]
	\centering
	\includegraphics[width=0.5\textwidth]{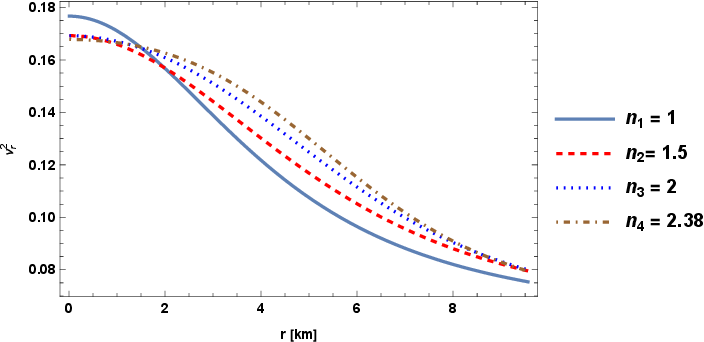} 
	\caption{
		\centering
		\textbf{Change in radial distribution of radial velocity ($\upsilon^2_r$)} \textbf{in {PSR J0952\textendash0607}} 
		\newline (\( M = 2.17 \, M_{\odot}; R = 9.56 \, \mathrm{km} \)) for various values of \( n \).
	}
	\label{fig:5}
\end{figure}

\begin{figure}[h!]
	\centering
	\includegraphics[width=0.5\textwidth]{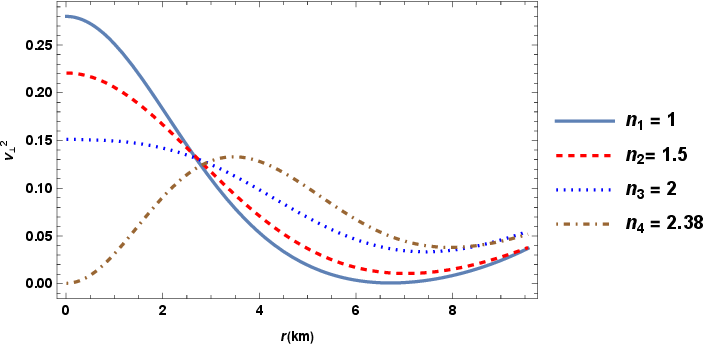} 
	\caption{
		\centering
		\textbf{Change in radial distribution of transverse velocity ($\upsilon^2_\perp$)} \textbf{in {PSR J0952\textendash0607} } 
		\newline (\( M = 2.17 \, M_{\odot}; R = 9.56 \, \mathrm{km} \)) for various values of \( n \).
	}
	\label{fig:6}
\end{figure}

\begin{figure}[h!]
	\centering
	\includegraphics[width=0.5\textwidth]{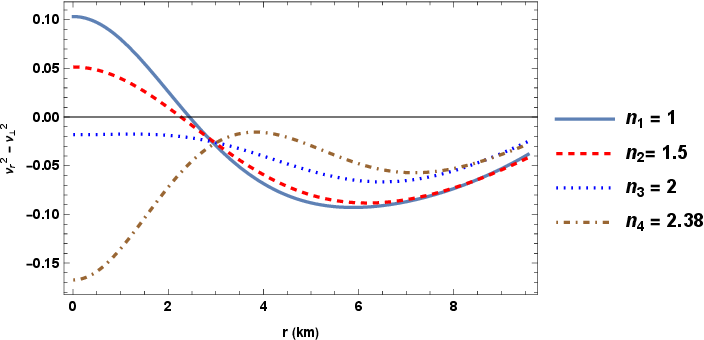} 
	\caption{
		\centering
		\textbf{change in radial distribution of ($\upsilon^2_\perp$- $\upsilon^2_r$)}  \textbf{in {PSR J0952\textendash0607} } 
		\newline (\( M = 2.17 \, M_{\odot}; R = 9.56 \, \mathrm{km} \)) for various values of \( n \).
	}
	\label{fig:7}
\end{figure}

Furthermore, Abreu's criteria  against Herrera's cracking approach are mathematically expressed as:
\begin{equation}
	\left| v_\perp^2 - v_r^2 \right| \leq 1 \implies
	\begin{cases} 
		-1 \leq v_\perp^2 - v_r^2 \leq 0 & : \text{Stable Region} \\
		0 < v_\perp^2 - v_r^2 \leq 1 & : \text{Unstable Region}  \label{equ:31}
	\end{cases}  
\end{equation}
The anisotropic model can be confirmed as stabilized by analyzing and plotting the radial $v_r^2$ and transverse $v_\perp^2$  sound velocities as depicted in Figure \ref{fig:5} and \ref{fig:6}, respectively. The analysis shows that the conditions $ 0 \leq v_r^2 \leq 1$ and $ 0 \leq v_t^2 \leq 1$ are achieved throughout the stellar system, validating the causality condition. According to another concept, a region where  $v_r^2$ is more than that $v_\perp^2$ is considered to be a stable distribution(\cite{abreu2007sound, chan1993dynamical, herrera1992cracking, herrera1979adiabatic}). To define a stable matter distribution, the stability criterion $\left| v_\perp^2 - v_r^2 \right| \leq 1 $ derived by Herrera and Andreasson is referred to as the "no cracking" condition. In Figure \ref{fig:7}, $v_\perp^2 - v_r^2$  for PSR J0952–0607 indicates that for \(n = 1\) and \(n = 1.5\), the value of $v_\perp^2 - v_r^2$ at the center is positive. The condition $v_\perp^2 - v_r^2 > 0$ suggests the possibility of cracking, which may cause the model to become unstable. For $-1 \leq v_\perp^2 - v_r^2 \leq 0$, the inequality condition is  achieved for $2 \leq n \leq 2.38$. Beyond this limit, a physically viable solution cannot be obtained graphically. The solutions up to \(n = 2.38\) demonstrate no cracking within the stellar interior, resulting in a stable stellar configuration \cite{abreu2007sound}.
\subsection{Adiabatic index}
The ratio of specific heats is measured by the adiabatic index $\Gamma$, which is important to understand EOS at different densities. It is essential for determining the stability of both relativistic and non-relativistic fluid spheres. Bondi \cite{bondi1964contraction} recognized the radial (\(\Gamma_r\)) and transverse (\(\Gamma_\perp\)) adiabatic indices in anisotropic fluid spheres as essential parameters for assessing their stability in response to infinitesimal radial adiabatic perturbations.\\ 
For a Newtonian sphere to be considered stable, $\Gamma$ must exceed $\frac{4}{3}$, and for neutral equilibrium, it is exactly $\frac{4}{3}$, as per Bondi's analysis \cite{bondi1964contraction}. In the relativistic context, the isotropic sphere's stability condition is altered due to regenerative pressure effects, potentially leading to significant instability. However, in a general relativistic, anisotropic sphere, the nature of anisotropy crucially influences stability, necessitating $\Gamma$ to exceed $\frac{4}{3}$ within a dynamically stable system, as supported by studies from Chan \cite{chan1993dynamical}, Heinzmann  \cite{heintzmann1975neutron}, and Hillebrandt \cite{hillebrandt1976anisotropic}. The model under consideration specifies the formulas for radial and transverse adiabatic indices as follows: 
\begin{equation}
	\Gamma_r = \bigg[\frac{\rho(r) + p_r(r)}{p_r(r)}\bigg] \bigg[\frac{dp_r}{d\rho}\bigg] = \bigg[\frac{\rho(r) + p_r(r)}{p_r(r)}\bigg]\big[v^2_{r}\big]
\end{equation}

\begin{equation}
	\Gamma_t = \bigg[\frac{\rho(r) + p_\perp(r)}{p_\perp(r)}\bigg] \bigg[\frac{dp_\perp}{d\rho}\bigg] 
	= \bigg[\frac{\rho(r) + p_\perp(r)}{p_\perp(r)}\bigg]\big[v^2_{\perp}\big] 
	\label{equ-33}
\end{equation}

\begin{figure}[h!]
	\centering
	\includegraphics[width=0.5\textwidth]{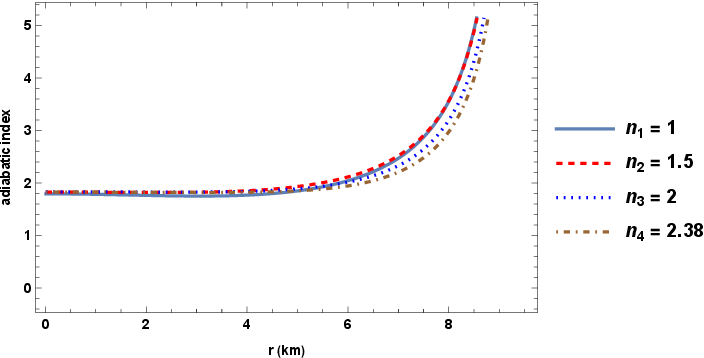} 
	\caption{
		\centering
		\textbf{change in radial distribution of ($\Gamma$)} \textbf{in {PSR J0952\textendash0607} } 
		\newline (\( M = 2.17 \, M_{\odot}; R = 9.56 \, \mathrm{km} \)) for various values of \( n \).
	}
	\label{fig:8}
\end{figure}

The graphical presentation of adiabatic indices is depicted in figure \ref{fig:8}. The adiabatic indices are greater than $\frac{4}{3}$, giving rise to stable stellar configuration against the radial perturbation.

\subsection{Energy conditions}
\begin{figure}[h!]
	\centering
	\includegraphics[width=0.5\textwidth]{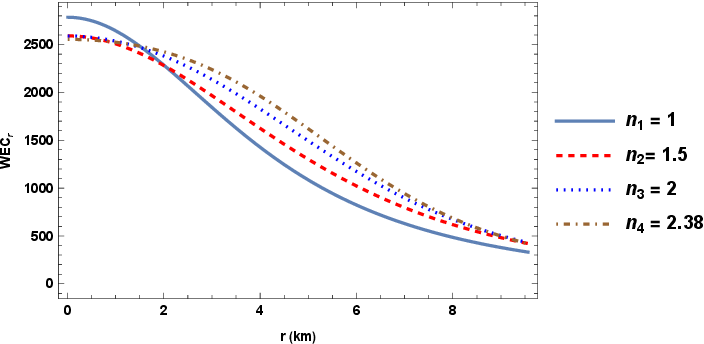} 
	\caption{
		\centering
		\textbf{Change in radial distribution of ($WEC_r$)} \textbf{in {PSR J0952\textendash0607} } 
		\newline (\( M = 2.17 \, M_{\odot}; R = 9.56 \, \mathrm{km} \)) for various values of \( n \).
	}
	\label{fig:9}
\end{figure}

\begin{figure}[h!]
	\centering
	\includegraphics[width=0.5\textwidth]{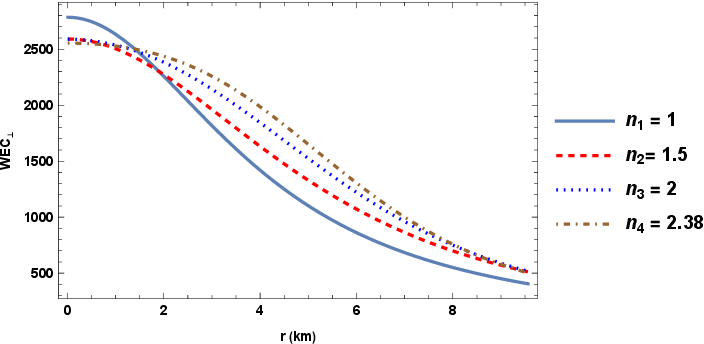} 
	\caption{
		\centering
		\textbf{Change in radial distribution of ($WEC_\perp$)} \textbf{in {PSR J0952\textendash0607} } 
		\newline (\( M = 2.17 \, M_{\odot}; R = 9.56 \, \mathrm{km} \)) for various values of \( n \).
	}
	\label{fig:10}
\end{figure}

\begin{figure}[h!]
	\centering
	\includegraphics[width=0.5\textwidth]{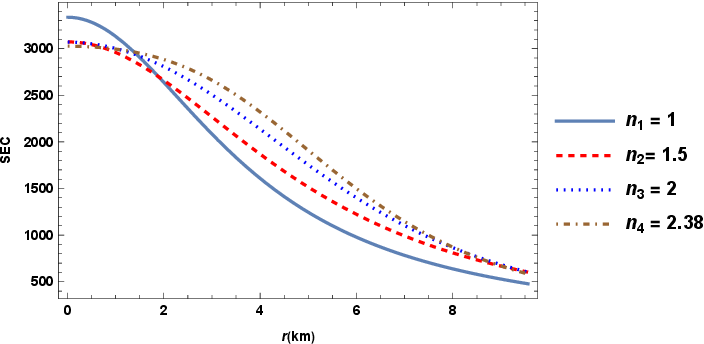} 
	\caption{
		\centering
		\textbf{Change in radial distribution of (SEC)} \textbf{in {PSR J0952\textendash0607} } 
		\newline (\( M = 2.17 \, M_{\odot}; R = 9.56 \, \mathrm{km} \)) for various values of \( n \).
	}
	\label{fig:11}
\end{figure}

\begin{figure}[h!]
	\centering
	\includegraphics[width=0.5\textwidth]{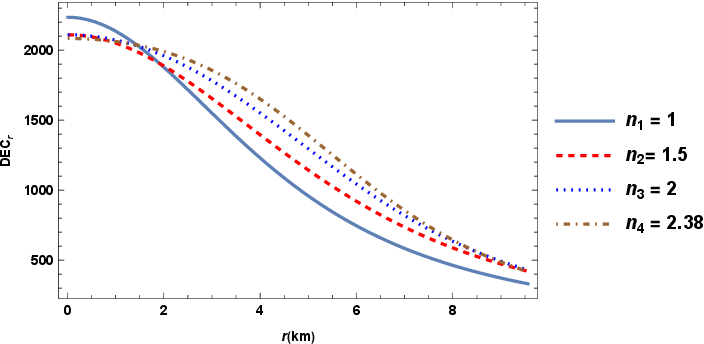} 
	\caption{
		\centering
		\textbf{Change in radial distribution of ($DEC_r$)} \textbf{in {PSR J0952\textendash0607} } 
		\newline (\( M = 2.17 \, M_{\odot}; R = 9.56 \, \mathrm{km} \)) for various values of \( n \).
	}
	\label{fig:12}
\end{figure}

\begin{figure}[h!]
	\centering
	\includegraphics[width=0.5\textwidth]{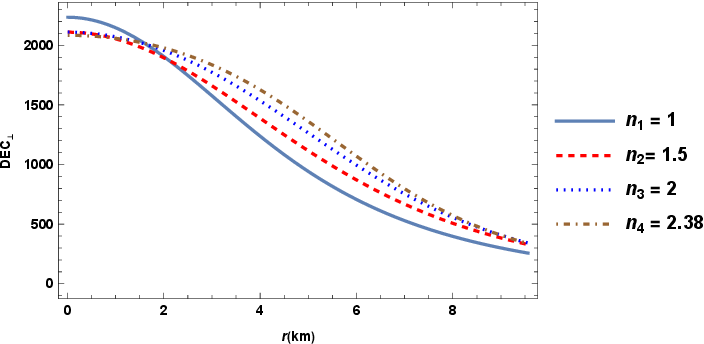} 
	\caption{
		\centering
		\textbf{Change in radial distribution of ($DEC_\perp$)}  \textbf{in {PSR J0952\textendash0607} } 
		\newline (\( M = 2.17 \, M_{\odot}; R = 9.56 \, \mathrm{km} \)) for various values of \( n \).
	}
	\label{fig:13}
\end{figure}

\begin{figure}[h!]
	\centering
	\includegraphics[width=0.5\textwidth]{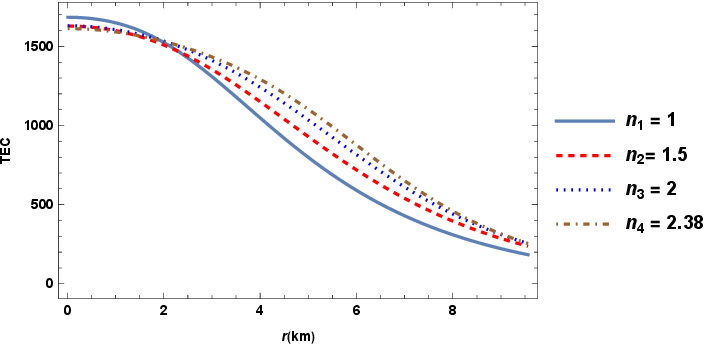} 
	\caption{
		\centering
		\textbf{Change in radial distribution of ($TEC$)} \textbf{in {PSR J0952\textendash0607} } 
		\newline (\( M = 2.17 \, M_{\odot}; R = 9.56 \, \mathrm{km} \)) for various values of \( n \).
	}
	\label{fig:14}
\end{figure}

The following are the energy conditions which are shown in figures \ref{fig:9} - \ref{fig:14}.

\begin{align}
	\text{NEC:} & \quad \rho \geq 0 \label{equ:34}  \\  
	\text{DEC}_r: & \quad \rho - |p_r| \geq 0 \label{equ:35} \\  
	\text{DEC}_\perp: & \quad \rho - |p_t| \geq 0 \label{equ:36} \\  
	\text{WEC}_r: & \quad \rho + p_r \geq 0 \label{equ:37} \\  
	\text{WEC}_\perp: & \quad \rho + p_t \geq 0 \label{equ:38} \\  
	\text{SEC:} & \quad \rho + p_r + 2p_\perp \geq 0 \label{equ:39} \\  
	\text{TEC:} & \quad \rho - p_r - 2p_\perp \geq 0 \label{equ:40}
\end{align}
The null energy condition (NEC) states that the energy density measured by an observer traveling at the speed of light is positive. Figures (\ref{fig:12}) and (\ref{fig:13}) show the graphical presentation of dominant energy condition in radial and transverse direction respectively. The dominant energy conditions $DEC_r$ and $DEC_\perp$ predicts that the density is always greater than the pressure. If it is violated then repulsive gravitational effects may arise leading to instability in compact object like neutron star. By verifying weak energy conditions $WEC_r$ and $WEC_\perp$, we ensure that the stellar model is stable and does not introduce the exotic matter. Figures (\ref{fig:9}) and (\ref{fig:10}) graphical representation of $WEC$.  The strong energy condition (SEC) ensures that the gravity always remains attractive. SEC ensures that the pressure and density contribute positive to make a stable stellar model. Figure (\ref{fig:11}) gives the notion that density and pressure are positive throughout the stellar distribution. Trace energy condition (TEC) in Figure (\ref{fig:14}) tells that energy-momentum tensor remains positive throughout the stellar interior.
So, it is noteworthy to see that all the energy conditions are satisfied inside the GTK metric system.


\subsection{Hydrostatic equilibrium under different forces}
The generalized Tolman-Oppenheimer-Volkoff (TOV) equation is used to analyze the model to assess the stability of a star under different forces as formulated by Ponce de León \cite{ponce1987general}, and is expressed as:
\begin{equation}
	-\frac{M_G (\rho + p_r)}{r^2}e^{\frac{\lambda-\nu}{2}}- \frac{dp_r}{dr} + \frac{2}{r}(p_\perp - p_r) = 0 \label{equ:41}
\end{equation}
where, $M_G = M_G(r)$ denotes effective gravitational mass enclosed within a sphere of radius r. This physical quantity can be obtained through the modified Tolman-Whittaker formula \cite{devitt1989modified}, interpreted as:

\begin{equation}
	M_G(r) = \frac{1}{2}r^2 e^{\left(\frac{\nu-\lambda}{2}\right)} \nu'. \label{equ:42}
\end{equation}
Substituting this value into equation (\ref{equ:36}) reformulates the TOV equation as follows:
\begin{equation}
	-\frac{(\rho + p_r)\nu'}{2}-\frac{dp_r}{dr}+ \frac{2}{r}(p_\perp - p_r) = 0. \label{equ:43}
\end{equation}
Equation (\ref{equ:43}) represents the equilibrium condition of the fluid sphere, considering the combined influence of hydrostatic, anisotropic and gravitational forces.
\begin{equation}
	F_g + F_h + F_a = 0. \space \label{equ:44}
\end{equation}
where,
\begin{align}
	F_g = & -A r \Bigg[ n (2a + 4b r^2) p + \frac{1 - s}{r^2} \nonumber \\
	& \quad + \Big( n (2a + 4b r^2) p + \frac{1 - s}{r^2} \Big)^2 \alpha \nonumber \\
	& \quad + \Big( n (2a + 4b r^2) p + \frac{1 - s}{r^2} \Big) \beta - \gamma \Bigg], \label{equ:45}
\end{align}

	\begin{multline}\label{equ:46}
	F_h = \frac{2}{r^5} q^{-3 - 2n} \biggl[ -1 + h + 2 b r^4 (-1 - 3 n + h) \\
	+ a^2 r^4 (-1 + n + 2 n^2 + h) + b^2 r^8 (-1 + 2 n + 8 n^2 + h) \\
	+ a r^2 \left(8 b n^2 r^4 + n (-1 + b r^4) + 2 (1 + b r^4)(-1 + h)\right) \biggr] \\
	\biggl[ 2 \left(-1 + h + a r^2 (-1 + 2 n + h) + b r^4 (-1 + 4 n + h)\right) \alpha \\
	+ r^2 q^{1+n} \beta \biggr].
\end{multline}

\begin{multline}\label{equ:47}
	F_a = \frac{1}{r^5}\biggl[-2 q^{-3 - 2n}\bigl(-1 + h + 2 b r^4(-1 - 3 n + h) \\
	+ a^2 r^4(-1 + n + 2 n^2 + h) + b^2 r^8(-1 + 2 n + 8 n^2 + h) \\
	+ a r^2\left(8 b n^2 r^4 + n(-1 + b r^4) + 2(1 + b r^4)(-1 + h)\right)\bigr)\biggr] \\
	\times \biggl[2\left(-1 + h + a r^2(-1 + 2 n + h) + b r^4(-1 + 4 n + h)\right)\alpha \\
	+ r^2 q^{1 + n}\beta\biggr] \\
	+ A r^2\biggl[2 n r^4(a + 2 b r^2)p + r^2(1 - s) \\
	+ \left(-1 - 2 n r^2(a + 2 b r^2)p + s\right)^2\alpha \\
	+ r^2\left(1 + 2 n r^2(a + 2 b r^2)p - s\right)\beta - r^4\gamma\biggr].
\end{multline}

where, 
\begin{align*}
	q &= 1 + a r^2 + b r^4, \quad s = q^{-n}, \quad p = q^{-1-n}, \quad h = q^n.
\end{align*}

\begin{figure}[h!]
	\centering
	\includegraphics[width=0.5\textwidth]{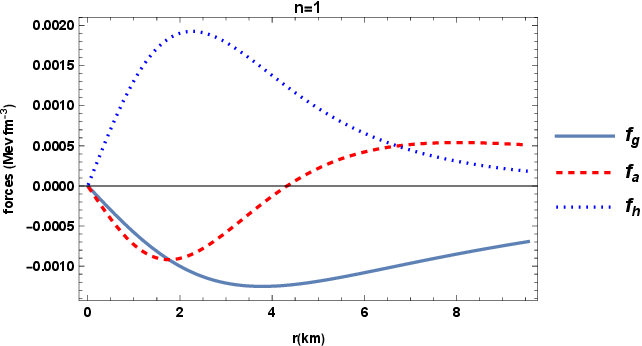} 
	\caption{
		\centering
		\textbf{Change in radial distribution of forces for (n = 1)} \textbf{in {PSR J0952\textendash0607} } 
		\newline (\( M = 2.17 \, M_{\odot}; R = 9.56 \, \mathrm{km} \)) for various values of \( n \).
	}
	\label{fig:15}
\end{figure}

\begin{figure}[h!]
	\centering
	\includegraphics[width=0.5\textwidth]{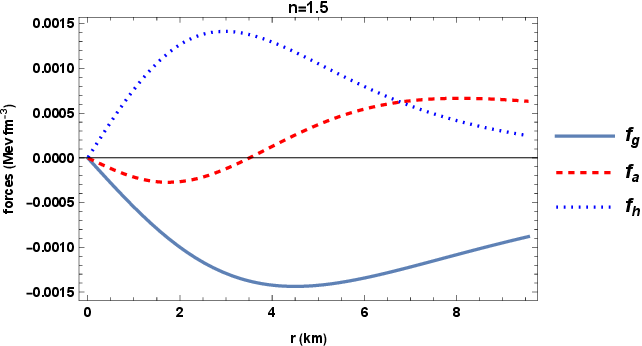} 
	\caption{
		\centering
		\textbf{Change in radial distribution of forces for (n =1.5)} \textbf{in {PSR J0952\textendash0607} } 
		\newline (\( M = 2.17 \, M_{\odot}; R = 9.56 \, \mathrm{km} \)) for various values of \( n \).
	}
	\label{fig:16}
\end{figure}

\begin{figure}[h!]
	\centering
	\includegraphics[width=0.5\textwidth]{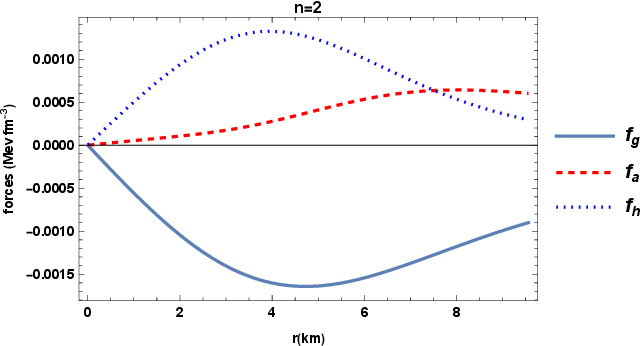} 
	\caption{
		\centering
		\textbf{Change in radial distribution of forces for (n = 2)} \textbf{in {PSR J0952\textendash0607} } 
		\newline (\( M = 2.17 \, M_{\odot}; R = 9.56 \, \mathrm{km} \)) for various values of \( n \).
	}
	\label{fig: 17}
\end{figure}

\begin{figure}[h!]
	\centering
	\includegraphics[width=0.5\textwidth]{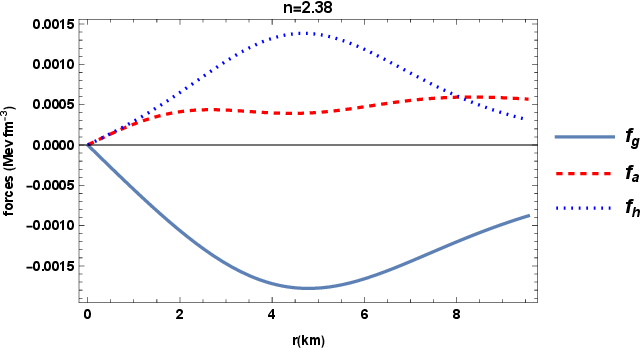} 
	\caption{
		\centering
		\textbf{Change in radial distribution of forces for (n = 2.38)} \textbf{in {PSR J0952\textendash0607} } 
		\newline (\( M = 2.17 \, M_{\odot}; R = 9.56 \, \mathrm{km} \)) for various values of \( n \).
	}
	\label{fig:18}
\end{figure}

Figure \ref{fig:15} - \ref{fig:18} show the fluctuations in the forces $F_g, F_h$, and $F_a$ for n = 1, 1.5, 2, and 2.38, respectively. The graph reveals that the gravitational force balances the combined effects of hydrostatic and anisotropic forces, keeping the ultra-dense compact stars in a stable state.
\subsection{Mass and compactness}
A relativistic star model is constructed based on its known mass and radius, with parameters chosen to ensure stable hydrostatic equilibrium. The mass of the millisecond pulsar is \(M_{NS} = 2.35 \pm 0.17 \, M_{\odot}\), and its radius is assumed to be 10 km. A previous study has shown that the lowest value for the maximum mass of a neutron star is \(M_{\text{max}} > 2.19 M_{\odot}\) (\(2.09 M_{\odot}\)) at \(1\sigma\) (\(3\sigma\)) confidence \cite{romani2022psr}. \(M = 2.17 M_{\odot}\) and a radius of 9.56 km are used to determine the model parameters for different \(n\) values, as shown in Table \ref{tab:Table 1}. It is observed that the star's central density exceeds that of its surface, and the model remains stable for \(2 \leq n \leq 2.38\).

\begin{figure}[h!]
	\centering
	\includegraphics[width=0.5\textwidth]{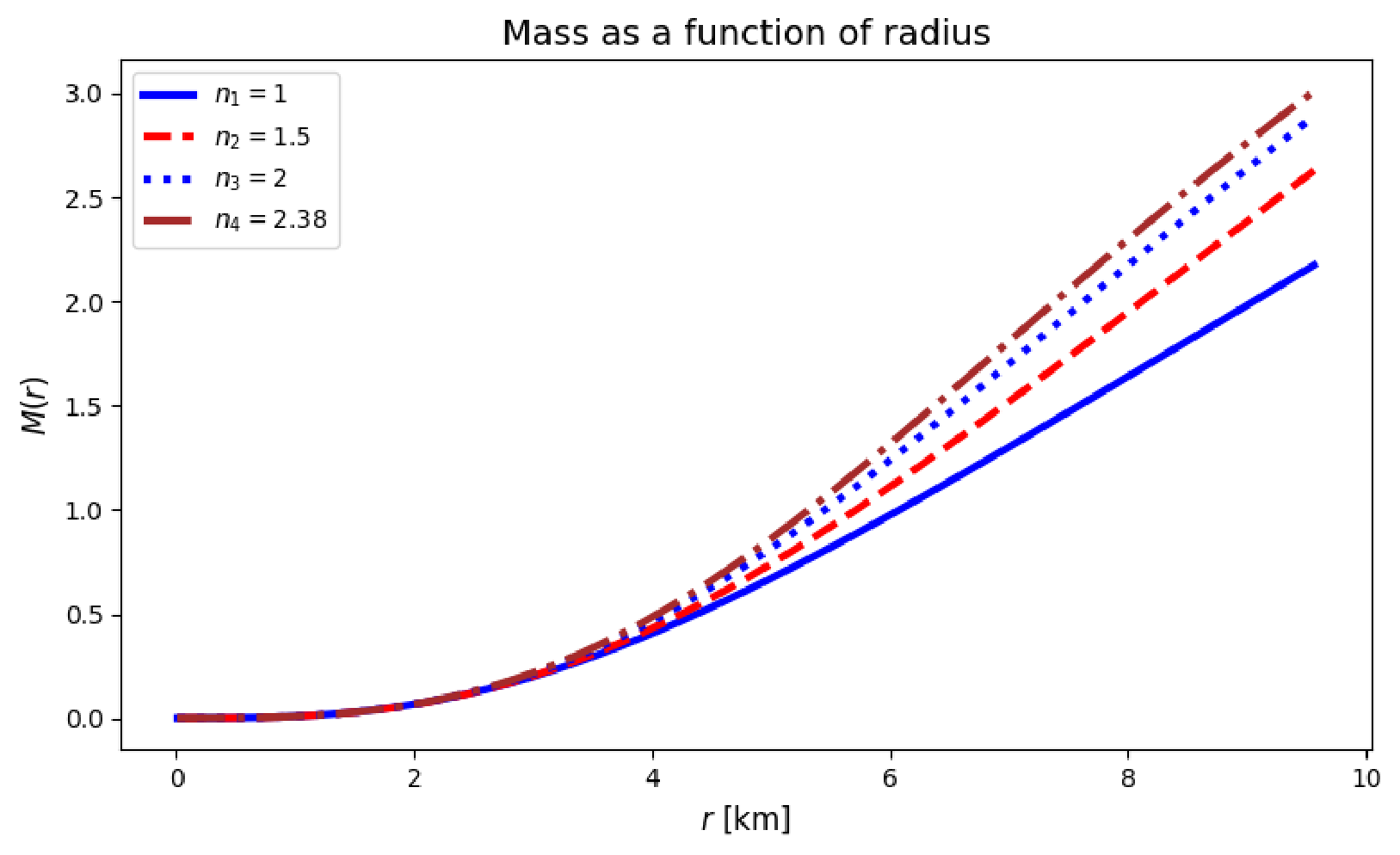} 
	\caption{
		\centering
		\textbf{change in radial distribution of mass} \textbf{in {PSR J0952\textendash0607}  \cite{das2024finch}} 
		\newline (\( M = 2.17 \, M_{\odot}; R = 9.56 \, \mathrm{km} \)) for various values of \( n \).
	}
	\label{fig:19}
\end{figure}

To validate the precision of any model, it is essential to ascertain the maximum mass and the mass-radius relationship. The effective gravitational mass in GR is determined by the distribution of an ideal fluid, whether it is charged or uncharged, anisotropic or isotropic.
\begin{equation}
	M_0(r) = 4\pi \int_0^r r^2 \rho(r) \, dr \label{equ:48}
\end{equation}
As depicted in Figure \ref{fig:19}, when \( r \to 0 \), the effective mass \( M_{\text{eff}} \) also approaches zero, illustrating its monotonic growth.

So, Buchdahl found the upper limit for the mass-radius ratio of an isotropic perfect fluid matter distribution, where the energy density decreases towards the boundary \cite{buchdahl1959general}.
\begin{equation}
	\frac{M_0}{R} \leq \frac{4}{9} \label{equ:49}
\end{equation}
R is the compact object's radius. Mak et al. \cite{mak2001maximum} generalized their findings to the charged spheres.
Stellar objects are classified based on their compactness \(\frac{M}{R}\).
\begin{itemize}
	\item (i) Neutron stars: \( 10^{-1} < M/R < 1/4 \),
	\item (ii) Ultra-dense stars: \( 1/4 < M/R < 1/2 \),
	\item (iii) White dwarfs: \( M/R \sim 10^{-3} \),
	\item (iv) Black holes: \( M/R = 1/2 \), and
	\item (v) Normal stars: \( M/R \sim 10^{-5} \).
\end{itemize}

\begin{figure}[h!]
	\centering
	\includegraphics[width=0.5\textwidth]{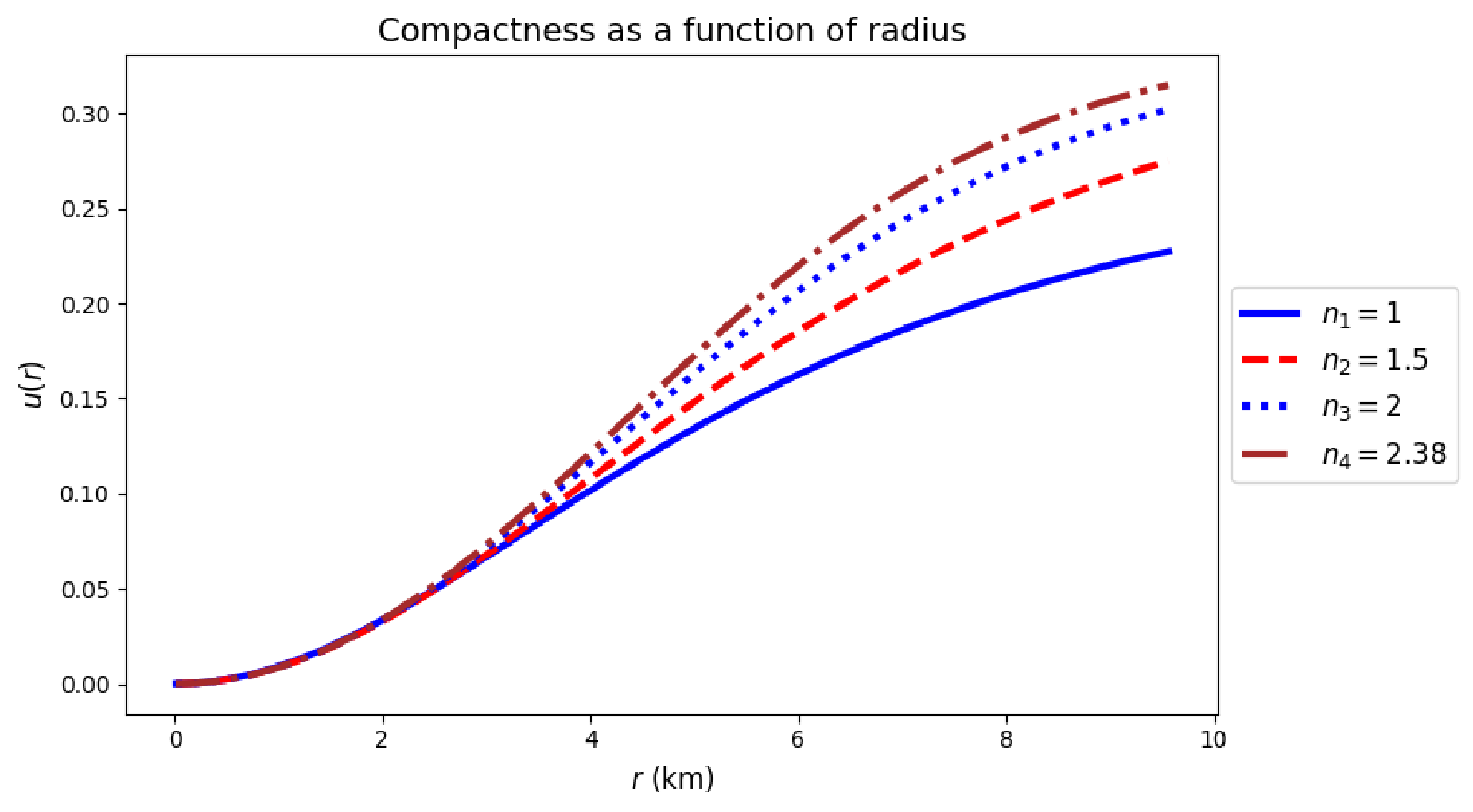} 
	\caption{
		\centering
		\textbf{Radial variation of compactness} \textbf{in {PSR J0952\textendash0607} } 
		\newline (\( M = 2.17 \, M_{\odot}; R = 9.56 \, \mathrm{km} \)) for various values of \( n \).
	}
	\label{fig:20}
\end{figure}

The ratio of the effective mass to the radius \(r\) is defined by the compactness \(u(r)\) of the model. It is mathematically represented as:
\begin{equation}
	u(r) = \frac{M_{\text{eff}}}{r} = \frac{1}{r} \int_0^R 4\pi r^2 \rho(r) \, dr. \label{equ:50}
\end{equation}
Compactness rises monotonically with the star's radius, as seen graphically in Figure \ref{fig:20}. The model represents an ultra-dense star if its maximum value is more than 0.25.

\section{Discussion}\label{sec7}

A relativistic model of an anisotropic uncharged compact star in Einstein's geometry is developed using the GTK metric. The GTK metric's exponent n is a critical factor in determining the matter configuration within the compact structure. The T-K metric (\(n = 1\)), is generalized to $n \geq 1$ to generate a physically feasible star model. The physical acceptability of PSR J0952\textendash0607 is investigated in this work by using its known mass and radius. Here, in our analysis, the range of n is taken as \(n \in [1, 2.38]\), since values of 
n beyond 2.38 violate the causality condition. We have analysed the energy density profile, pressure profiles, and all other physical parameters with the increasing values of n. Identifying the precise solutions for confined objects is difficult due to the complexities and non-linearity of relativistic field equations. Therefore, numerical methods have been employed.\\
These are the major findings of the present study:
\par The energy density and pressure profiles are positive and satisfy all the physical criteria as seen in Figures \ref{fig:1}–\ref{fig:3}, respectively. Figure \ref{fig:1} shows that as the values of \(n\) increase, the energy density decreases. The central energy density for \(n\) in the range \(n \in [1, 2.38]\) is determined to fit within the limit of \(2510 - 2319 \,\mathrm{MeV} \,\mathrm{fm}^{-3}\), while the density at the surface is confined to the range \( 332 - 420 \, \mathrm{MeV} \,\mathrm{fm}^{-3} \).

Figures \ref{fig:2} and \ref{fig:3} state that, the radial pressure (\(p_r\)) reduces to zero at the boundary. On the contrary, the transverse pressure (\(p_t\)) decreases as \(n\) increases but does not vanish at the boundary. At the center of the star, the value of radial and transverse pressures are equal.
\par As seen in Figure \ref{fig:4}, the anisotropy is zero at the center. For \(n = 1\) and \(n = 1.5\), the radial pressure is higher than that of the transverse pressure giving rise to the inward force causing the instability in the model. However, when \(n > 1.5\), the transverse pressure exceeds the radial pressure, resulting in a stable configuration for the PSR J0952\textendash0607. Thus, the stable structure of the mathematical model is achieved. 
\par The causality conditions and Herrera's cracking condition have been represented in Figures \ref{fig:5} - \ref{fig:7}. The speed of sound has been measured to determine the stability of the stellar structure. The radial speed of sound, \(v_r^2\), decreases as \(n\) increases, but the transverse speed of sound, \(v_\perp^2\), decreases more abruptly with increase in n. Graphically, the transverse sound velocity is not within the range (\(0 \leq v_\perp^2 \leq 1\)) beyond the value of \(n = 2.38\). When $n = 1$ and $1.5$, the quantity $v_{\perp}^{2} - v_{r}^{2}$ is positive at the core, causing instability in the star. Nevertheless, the condition $-1 \leq v_{\perp}^{2} - v_{r}^{2} \leq 0$ holds true up to $n = 2.38$, ensuring stable stellar configurations within this range.

\par The adiabatic index in Figure \ref{fig:8} illustrates the EoS's stiffness. The adiabatic index is greater than \(\frac{4}{3}\) for relativistic anisotropic stars, suggesting that a stable configuration has been obtained.
\par Figures \ref{fig:9} – \ref{fig:14} represent the radial variations of several energy conditions, including the dominant, trace, strong, weak, and null energy conditions. The uncharged compact stellar model for PSR J0952\textendash0607 adheres to each energy condition when the GTK metric is employed. The gradients of various forces are shown in figures \ref{fig:15} –  \ref{fig:18}, which include the gradients of the gravitational force (\(F_g\)), the hydrostatic force (\(F_h\)), and the anisotropic force (\(F_a\)) for various values of \(n\). These graphs indicate that \(F_g\) is always positive, \(F_h\) is always negative, and \(F_a\) exhibits mixed behaviour, all contributing to a stable configuration.
\par Figures \ref{fig:19} and \ref{fig:20}  depict the M-R relationship and compactness factor, respectively. It is evident from the figure that the M-R relation and compactness factor increase as n increases. Thus, the present investigation fulfills Buchdahl's condition.
By employing the GTK metric ansatz, we have obtained solutions for a static, spherically symmetric matter distribution using the QEoS. When \( n = 1 \) and \( n = 1.5 \), the inequality  
$
-1 \leq v_{\perp}^2 - v_r^2 \leq 0
$ 
is not satisfied. Stability is achieved only for \( n > 1.5 \). However, beyond \( n = 2.38 \), the square of the transverse sound velocity moves outside the permissible range of 0 to 1, as obtained from the graphical presentation. Therefore, the stable structure of the anisotropic matter distribution extends up to \( n = 2.38 \).

\section*{Abbreviations}

\begin{description}[align=left, labelwidth=1.60cm, itemsep=0.0001em]
	\item[GR] General Relativity 
	\item[EFEs] Einstein Field Equations 
	\item[EoS] Equation of State 
	\item[NS] Neutron Stars 
	\item[BH] Black Holes 
	\item[T-K] Tolman-Kuchowicz  
	\item[GTK] Generalized Tolman-Kuchowicz 
	\item[QEoS] Quadratic EoS 
	\item[TOV] Tolman-Oppenheimer-Volkoff
\end{description}

\section*{Declarations}
All authors have read, understood, and have complied as applicable with the statement on \enquote{Ethical responsibilities of Authors} as found in the Instructions for Authors.

\section*{Acknowledgments}
Ms. Hemani R. Acharya is grateful for the financial assistance provided by Pandit Deendayal Energy University, Gandhinagar, Gujarat, India, in the form of a fellowship.

\section*{Declaration of generative AI in scientific writing:} 
During the preparation of this work, the authors used Quillbot to improve readability and language. After using this tool, the authors reviewed and edited the content as needed and took full responsibility for the content of the publication.

\section*{Author contribution:} 
All the authors make a substantial contribution to this manuscript. All the authors discussed the results and implications of the manuscript at all stages. \textbf{Hemani R. Acharya:} Conceptualization, Formal analysis, Methodology, Software, Writing-original draft, Writing – review \& editing. \textbf{Dishant M. Pandya:} Supervision, Software, Resources, Validation, Investigation, Writing – review \& editing. \textbf{Bharat Parekh:} Supervision, Validation, Project administration, Writing – review \& editing.

\section*{Conflict of interest/Competing interests:} 
The authors have no relevant financial or non-financial interests to disclose.

\section*{Funding:} 
The authors declare that no funds, grants, or other support were received during the preparation of this manuscript.

\section*{Consent for publication:} 
Not applicable.

\section*{Ethical approval:} 
Not applicable.

\section*{Consent to participate:} 
Not applicable.

\section*{Data availability:} 
Not applicable.

\bibliographystyle{unsrt}
\bibliography{main}

\end{document}